\documentclass[final,3p,times]{elsarticle}

\usepackage[utf8]{inputenc} 
\usepackage{url}            
\usepackage{amsfonts}       
\usepackage{nicefrac}       
\usepackage{microtype}      
\usepackage{lipsum}
\usepackage{graphicx,floatrow,multirow}
\usepackage{xcolor}
\usepackage{amsmath,amssymb}
\usepackage{amsmath}
\usepackage{lineno,hyperref}
\usepackage{placeins}
\usepackage{graphicx,amssymb,amsmath,color,float,mathtools}
\usepackage{enumitem}
\usepackage{soul}
\usepackage{inputenc}
\usepackage{blindtext}
\usepackage{booktabs,multirow}
\usepackage{color,soul}
\usepackage{amsthm,mathrsfs,amsfonts,bm}
\usepackage[british]{babel}
\usepackage{caption}
\usepackage[export]{adjustbox}
\usepackage{subfig}
\usepackage{mwe}

\begin{document}
	
	\begin{frontmatter}
\title{Memory effects on link formation in temporal networks: A fractional calculus approach}
\author[1]{Fereshteh Rabbani}

\address[1]{Department of Physics, Shahid Beheshti University G.C., Evin, Tehran 19839, Iran}
\author[2]{Tamer Khraisha}
\address[2]{Department of Network and Data Science, Central European University, Nador U. 9, H-1051 Budapest, Hungary}

\author[1]{Fatemeh abbasi}
\cortext[cor1]{Corresponding author}

\author{Gholam Reza Jafari\corref{cor1}\fnref{1,2}}
\ead{g\_jafari@sbu.ac.ir}

\begin{abstract}
	Memory plays a vital role in the temporal evolution of interactions of complex systems. To address the impact of memory on the temporal pattern of networks, we propose a simple preferential connection model, in which nodes have a preferential tendency to establish links with most active nodes. Node activity is measured by the number of links a node observes in a given time interval. Memory is investigated using a time fractional order derivative equation, which has proven to be a powerful method to understand phenomena with long-term memory. The memoryless case reveals a characteristic time where node activity behaves differently below and above it. We also observe that  dense  temporal  networks (high  number  of  events)  show a  clearer characteristic time  than  sparse  ones. Interestingly, we also find that memory leads to decay of the node activity; thus, the chances of a node to receive new connections reduce with the node's age. Finally, we discuss the statistical properties of the networks for various memory-length.
\end{abstract}


\end{frontmatter}

\section*{Introduction}

Analyzing complex systems as compositions of entities and their interactions using network theory is a main trend in mathematical physics ~\cite{clauset2009power,newman2011structure,latora2017complex}. 
In the real world, systems often have the self-dynamic structure; therefore, they can be better described as networks in which links among a fixed set of nodes change over time ~\cite{holme2012temporal,masuda2016guidance}.
For example, activities like communication through social media ~\cite{holme2005network}, market trading ~\cite{halinen2013network}, and online search often take place in time ~\cite{li2017fundamental}. To understand the temporal dimension of these phenomena, the concept of temporal networks has been employed~\cite{kostakos2009temporal,kim2012temporal,wang2017link,hassanibesheli2017glassy}.
In this regard, it has been found that a non-Markovian dynamics is necessary to capture the non-trivial temporal patterns of real-world networks ~\cite{kim2015scaling,scholtes2014causality,vestergaard2014memory}, and can play an important role in processes occurring on temporal networks. 



An important aspect in temporal networks concerns the complex behavior  of agents, i.e. the  intuition  behind  each agent's decision  to  initiate connections toward  certain agents as well as the  strength  of established links. Individuals remember their social circle of friends and are likely to perform repeated interactions within their established circle. In other words, the formation of links can be though to be a Markov process ~\cite{pfitzner2013betweenness,karsai2014time,scholtes2013slow,karsai2012correlated,lambiotte2014effect,williams2019effects}. While the impact of memory has been studied in detail in static networks ~\cite{granovetter1977strength,dodds2003experimental,onnela2007structure},
little has been devoted to study the non-Markovian aspects in the context of temporal networks. For this reason, we conduct this study by focusing on long-range memory effects, where an arbitrary length of node memory can be included in the modeling of temporal network evolution. 
For  a  detailed discussion  on  the  origin  of memory effects, several attempts have been performed, where memory is considered as exploration ~\cite{karsai2014time},  preferential  return ~\cite{song2010modelling}, and social  reinforcement ~\cite{karsai2012correlated}.


\begin{figure}[!h]
    \centering
    \includegraphics[width=7cm , height=6cm ]{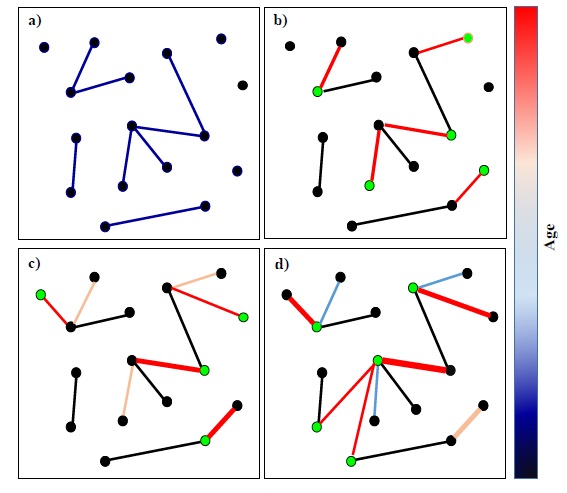}
    \caption{Illustration of a simple time-varying network by considering aging effect in the dynamics (a-d). At each instant,  $m$ (here, $m = 5$) nodes (green circles) are randomly selected and linked to the rest of nodes based on the node activity rule. 
        Red to blue color shades of links represent the age. Fresh links are in red.}\label{fig0}
\end{figure}

The field of \textit{fractional calculus} represents a promising solution to model memory effects of dynamic systems~\cite{metzler2000random,richard2014fractional,ebadi2016effect,saeedian2017memory,safdari2016fractional}, and has been recently used in the study of temporal networks ~\cite{berkowitz2002physical,kaslik2012nonlinear,west2015fractional}. 
Typically, the temporal dynamics of node activity is described with a differential equation, with the derivative being of integer order. When we calculate the value of an integer-order derivative at a specific point, the result depends only on that point. This property is called \textit{locality}. By replacing the ordinary time derivative with a fractional derivative, a time correlation function or memory kernel can be used, thereby making the state of the system dependent on all past states. This property is called\textit{nonlocality}, which is the built-in feature of fractional derivatives that allows them to capture memory effects.  For this reason, methods based on derivatives with non-integer
order, as introduced by Caputo for geophysics problems ~\cite{Caputo}, constitute an important formalism for non-Markovian problems. Moreover, Caputo's formalism provides the advantage that it is not necessary to define the fractional-order initial conditions when solving such differential equations ~\cite{Caputo,podlubny,podlubny2001geometric}.
Furthermore, the time correlation function, in the definition of Caputo fractional derivative, is a power-law function, which is flexible enough to reflect the fact that the contribution of earlier states on the current state is noticeably less relevant than the contribution of more recent states.

Using fractional calculus, this paper proposes a modified preferential attachment model by incorporating memory effects into the dynamics. The rest of the paper is structured as follows. First, we offer a definition of node activity and memory in our model. Next, relying on Caputo's approach, we convert the differential equation of our model to a fractional-order derivative to integrate memory. 
The main result of our work is to show that memory can play either of two opposing roles: it can slow-down or speed-up node activity depending on the strength of memory, which could be used as an alternative explanation of the findings of many other studies ~\cite{scholtes2014causality,karsai2014time,kim2015scaling,safdari2016fractional}. Fig.\ref{fig0} schematically illustrates a simple time-varying network considering aging effects in the dynamics. At each instant, we randomly choose $m$ nodes in the network and connect them with the rest of the nodes. A new edge is more likely to be established between a chosen node and the most active node(s). This illustration shows that the aging of links over time decreases the chance of high-degree nodes to receive new links.

\section{Modeling temporal networks}

Generating temporal networks is conceptually simple and work has been done in this area ~\cite{grindrod2009evolving,holme2012temporal,williams2019effects}. Significant attention has been devoted to temporal networks generated by the interactions of individuals in agent-based models ~\cite{buscarino2008disease,perra2012activity,starnini2013modeling,karsai2014time}. 
In the simplest setting, the researchers start with a set of nodes that connect with each other with time following a specific rule.


\begin{figure}
    \centering
    \subfloat[Numerical]{\label{a}\includegraphics[width=.5\linewidth]{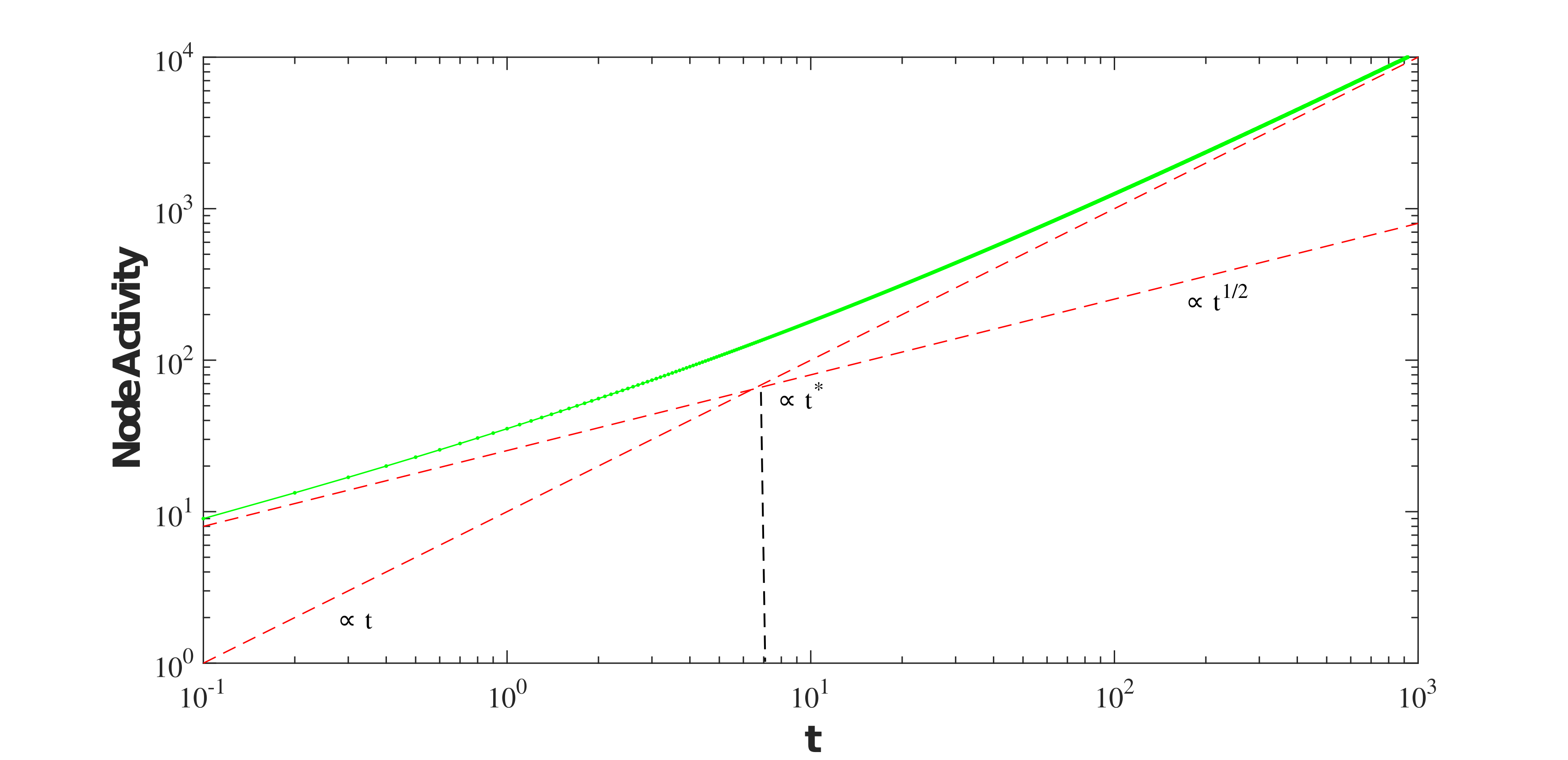}}\hfill
    \subfloat[Simulation]{\label{b}\includegraphics[width=.5\linewidth]{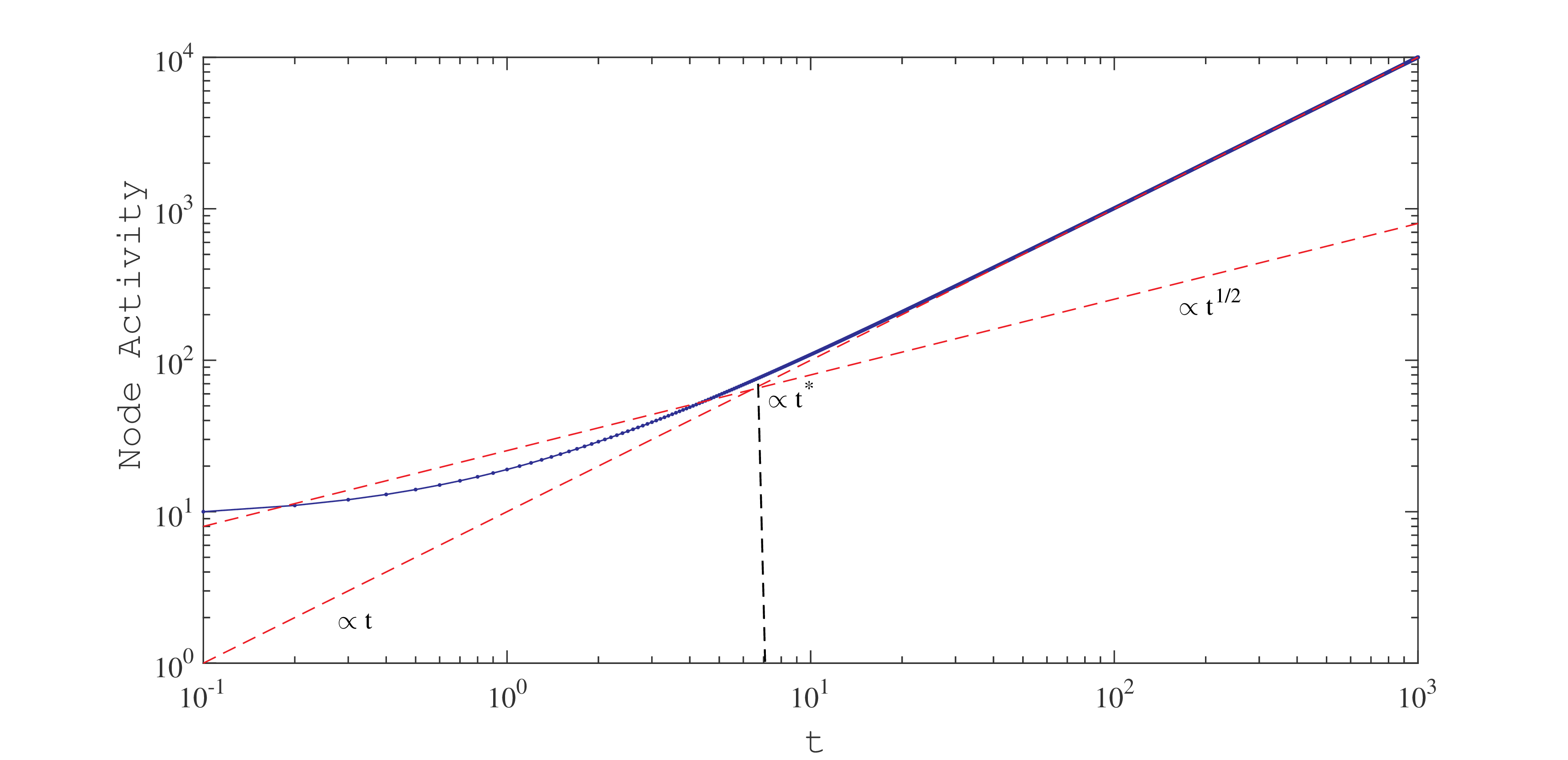}}\par 
    \subfloat[]{\label{c}\includegraphics[width=.45\linewidth]{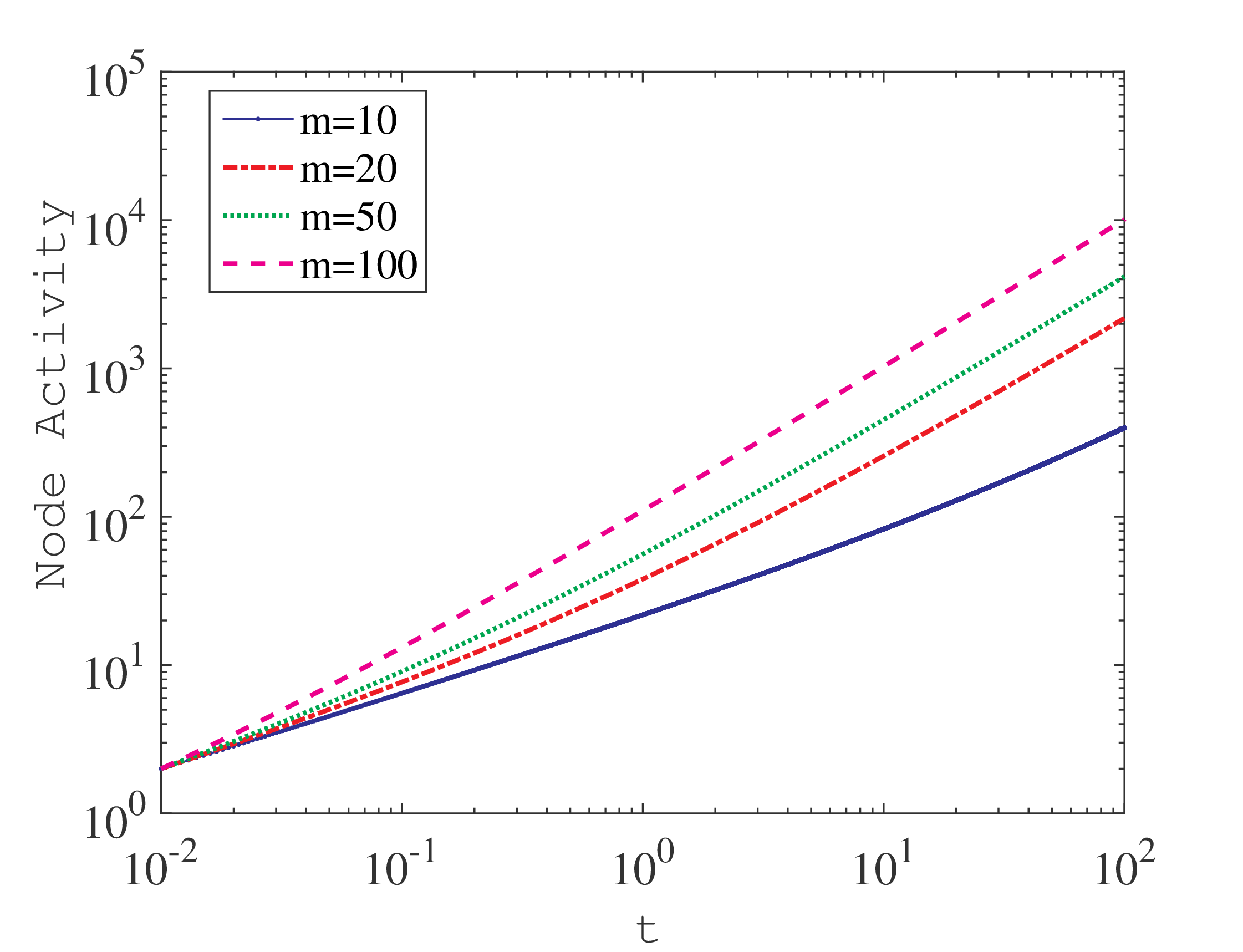}}
    \caption{(a) The analytical solution of Eq.\ref{eq:4} for the node activity $k_{i}(t)$ and (b) the corresponding result from network
    	simulation. Such simulation has been performed
    	for $t=N$ time steps, with an initial condition $N=1000$ nodes and $m = 10$. Here, $t^{*}=\frac{c^{2}N^{2}}{4m}$ is a characteristic time of the cross over to change the activity treat. (c) Node activity dynamics for different values of $m$.}
    \label{fig1}
\end{figure}

In what follows, we define a simple model for generating temporal networks based on the preferential attachment mechanism ~\cite{barabasi1999emergence,barabasi1999mean}.
The hypothesis behind our model is that each node prefers to create a link with the most active nodes. Node activity is defined as the number of events in which a node participates. Active nodes in a given time slot $\delta t$ participate in the largest number of the events that occur in a specific time window. Such nodes play a key role in the formation of the network topology as well as in the dissemination of information.

Let $L_{i}(t)$ be the number of events in which node $i$ appeared in the period from $t$ to $t + \Delta t$. We consider a network with $N$ nodes and assign to each node $i$ an activity rate $k_{i}(t) = \sum_{t=0}^{t} L_{i}(t)$. The activity rate is the probability per unit time to create new contacts with other individuals (i.e. the sum of temporal events from $t = 0$ to $t$). At each discrete time step $t$,  we randomly choose $m$  nodes who will establish an event with a target node based on the target node’s activity rate. This model is random and Markovian in the sense that agents do not have a memory of previous time steps. The full dynamics of the network and its ensuing structure is thus completely encoded in the node activity distribution $k_{i}(t)$. Thus, in mathematical terms, the activity of node $i$ evolves as:

\begin{equation}\label{eq:1}
\begin{aligned}
\frac{dk_{i}(t)}{dt} = \frac{m}{N}+\frac{mk_{i}(t)}{\sum_{j=1}^{N} k_{j}(t)}
\end{aligned}
\end{equation}

The first term on the right side of Eq.\eqref{eq:1} represents the percentage of nodes chosen at random to engage in new events. The second term adds a preferential connection component which we assume to be proportional to node activity. 
The model enables us to perform simple analytical calculations. 
The instantaneous network generated at each time $t$ will be composed of those nodes that were chosen randomly at that particular time, plus those who received connections from the randomly selected nodes. Since the total node activity of the network will increase over time, the total number of events will also increase as a function of $t$. By observing that each time $m$ new events are formed, and since each event has two ends, the sum of node activity increases by $2m$ at each time step. By Substituting this into \eqref{eq:1}, we obtain an analytical solution of the form:

\begin{equation}\label{eq:4}
\begin{aligned}
k_{i}(t)= \frac{2mt}{N}+c \sqrt{t}
\end{aligned}
\end{equation}

where  $c$  is  a  constant value and  depends  on  the  initial  conditions. At time $t=t_{0}$, the initial node activity is $k(t  = t_{0})=k_{0}$ and so  $c = \frac{k_{0}-2mt_{0}/N}{\sqrt{t_{0}}}$. Such results can be categorized into two classes of the activity as follows:\\

\begin{equation}
k_{i}(t) \sim
\begin{cases}
t, & \text{if}\  t \to \infty \\
\sqrt{t}, & \text{if}\ t \to 0
\end{cases}
\end{equation}

To validate Eq.~\eqref{eq:4} through numerical simulation, we examined a random network with $N=1000$ nodes and simulated node activities using Eq.~\eqref{eq:1}(Fig.~\ref{fig1}-(a)). At each time step, a total number of $m$ events are added where a source node $i$ is chosen at random to create a link with a target node $j$ with probability proportional to $k_j$. These steps are repeated sequentially, creating a network that grows with the passage of time. 
Fig.\ref{fig1}-(a) shows the node activity per time for both the analytical calculations and the ensemble average over realizations. These observations are in agreement with the BA model~\cite{barabasi1999mean}. The intersection of $t^{1/2}$ and $t$ is introduced as a characteristic time $t^{*}=\frac{c^{2}N^{2}}{4m^{2}}$ to detect behavior change. Fig.~\ref{fig1}-(b) plots the node activity dynamics for various $m$ in Eq.~\eqref{eq:4}. As we can see, for smaller values of $m$, the characteristic time is delayed.

\section{Memory effects on node's activity}

Next we extend the modeling framework by introducing memory. Conceptually, the idea of momory comes into play when nodes prefer to connect not only to the most active nodes, but also to those nodes with whom they had interactions in the past. In order to observe the effect of memory, first we rewrite the differential Eq.~\eqref{eq:1} in terms of time-dependent integrals as follows: 

\begin{equation}\label{eq:6}
\begin{aligned}
\frac{dk_{i}(t)}{dt}=\int_{t_0}^t \kappa (t-t^{'})dt^{'} [\frac{m}{N}+\frac{mk_{i}(t^{'})}{\sum_{j=1}^{N} k_{j}(t^{'})}]
\end{aligned}
\end{equation}

where $\kappa(t-t^{'})$ represents a time-dependent kernel and is equal to a delta function $\delta (t-t^{'})$ in a classical Markov
process. In fact, any arbitrary function can be replaced by a
sum of delta functions, thereby leading to some sort of time
correlation. A proper choice to include long-term
memory effects can be a power-law function, which exhibits
a slow decay such that the state of the system at quite early
times also contributes to the evolution of system. This type
of kernel guarantees the existence of scaling features as it is
often intrinsic in most natural phenomena. Thus, we consider the power-law correlation function for $\kappa(t -t^{'})=\frac{(t-t^{'})^{\alpha-2}}{\Gamma (\alpha-1)}$  where $0 \leqslant \alpha<1$ and $\Gamma$ denotes the Gamma function.



\begin{figure*}
    \centering
    \subfloat[Numerical]{{\includegraphics[width=6.3cm, height=5.5cm]{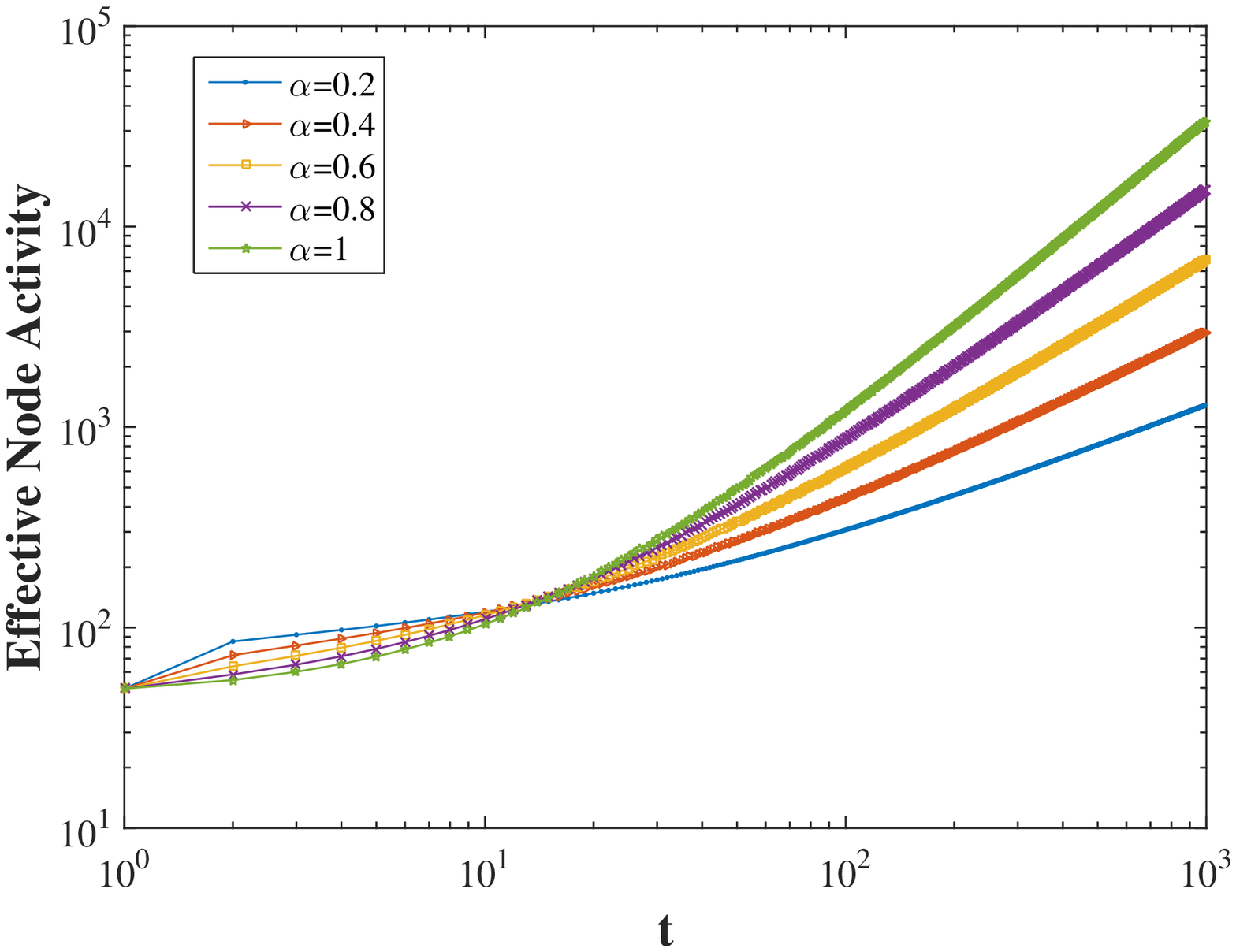}}}%
    \qquad
    \subfloat[Simulation]{{\includegraphics[width=6.4cm,height=5.5 cm]{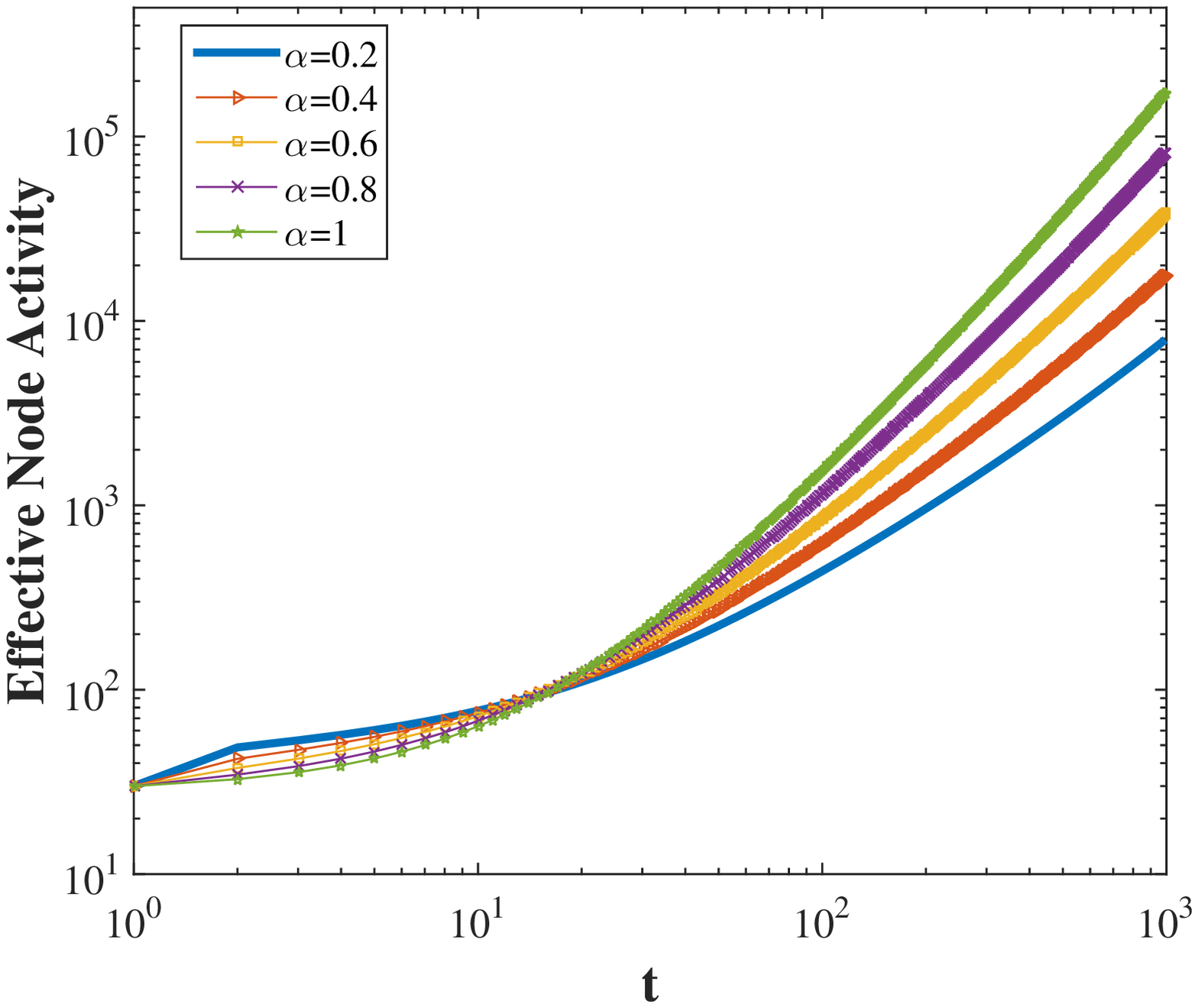}}}%
    \caption{(a)The numerical solution of Eq.~\ref{eq:11} for the effective node activity;  (b)The corresponding effective node's activity from network
        simulation, when its dynamics is affected by memory. It is clear from such graphs that $\bar k_{i}(t)$ behaves differently from the unlimited growth
        defined by the connection model $(\alpha =1)$. Due to the aging process $\bar k_{i}(t)$ reaches a peak and declines gradually afterwards. Such simulations have been performed for $t=1000$ time steps, with as initial condition $m= 10$ nodes and every new node connecting to earlier nodes.}%
    \label{fig2}%
\end{figure*}

The choice of the coefficient $\Gamma (\alpha-1)$ and exponent $(\alpha-2)$ allows us to rewrite Eq.~\eqref{eq:6} in the form of  a Caputo-like fractional differential equation.  By substituting this kernel function in Eq.~\eqref{eq:6}, we obtain a fractional integral of order $(\alpha-1)$ on the interval $[t_{0},t]$, denoted by ${}^{c}_{t_{0}}D^{-(\alpha-1)}_{t}$. Thus, we arrive to the following fractional differential equation:

\begin{equation}\label{eq:7}
\begin{aligned}
\frac{dk_{i}(t)}{dt}={}^{c}_{t_{0}}D^{-(\alpha-1)}_{t}[\frac{m}{N}+ \frac{mk_{i}(t^{'})}{\sum_{j=1}^{N} k_{j}(t^{'})}]
\end{aligned}
\end{equation}

If we set $\alpha = 1$, then the fractional operator turns to unity ${}^{c}_{t_{0}}D^{0}_{t}=1$ and Eq.~\eqref{eq:7} becomes identical to our model without memory described by Eq.~\eqref{eq:1}. Applying a fractional Caputo derivative of order $\alpha-1$ on both sides of
the above equation, we can write it in the form of a differential equation,

\begin{equation}\label{eq:8}
\begin{aligned}
{}^{c}_{t_{0}}D^{\alpha}_{t}(k_{i}(t)) =[\frac{m}{N}+ \frac{mk_{i}(t)}{\sum_{j=1}^{N} k_{j}(t)}]
\end{aligned}
\end{equation}

where ${}^{c}_{t_{0}}D^{\alpha}_{t}$ defined for an arbitrary function $y(t)$ as follows ~\cite{Caputo}:

\begin{equation}\label{eq:8-1}
\begin{aligned}
{}^{c}_{t_{0}}D^{\alpha}_{t}y(t)=\frac{1}{\Gamma (1-\alpha)} \int_{t_{0}}^{t} (t-s)^{-\alpha}y(s) ds
\end{aligned}
\end{equation}

Therefore, the fractional derivatives, when introducing a convolution
integral with a power-law memory kernel, are useful to
describe memory effects in dynamical systems. The decaying
rate of the memory kernel (a time-correlation function)
depends on $\alpha$. The upper value of $\alpha$ corresponds to a slowly
decaying time-correlation function (long memory). Hence,  the strength (through the "length") of the memory is controlled by $\alpha$. As $\alpha \to 1$, the effect of memory decreases: the system tends toward a memoryless system. 

Let's reemphasize that Eq.~\eqref{eq:8}, in the following general form:

\begin{equation}\label{eq:8-2}
\begin{aligned}
&{}^{c}_{t_{0}}D^{\alpha}_{t}(k_{i}(t))=f(t,k_{i}(t))\\&
k_{i}(t_{0})=k_{i0}
\end{aligned}
\end{equation}

Where $k_{i0}$ is a constant value that indicates the initial conditions. 
To deal with this equation, we use the predictor-corrector algorithm ~\cite{diethelm,garrappa2010linear}. It is assumed that
there exits a unique solution for  $k_{i}$ on the interval
$[0,t]$ for a given set of initial conditions. Considering a
uniform grid $\{t_{j}= t_{0}+jh : j = 0,1,2,..,N\}$ with equal space $h$, in which $N$ is
an integer.  Finally, the Eq.\eqref{eq:8-2} can be rewritten in a discrete form as follows:

\begin{figure}
    \hspace*{-1cm}
    \centering
    \includegraphics[width=6cm , height=3.50 cm ]{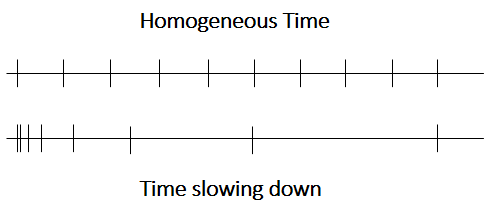}
    \caption{Schematic representation of fractional vs homogeneous time axis:  in the homogeneous time order, all units of time have the same length, while in the fractional order, time units have increasing length.}\label{fig4}
\end{figure}

\begin{equation}\label{eq:11}
\begin{aligned}
\bar k_{i}(t_{n}) = k_{i0} + h^{\alpha} \sum_{j=0}^{n-1} b_{n-j-1}f(t_{j},k_{j}(t))
\end{aligned}
\end{equation}

Here, $\bar k_{i}(t)$ denotes the effective node activity and $b_{n}$'s are time-dependent coefficients which represent the aging effect and they are equal to $b_{n}=\frac{(n+1)^{\alpha}-n^{\alpha}}{\Gamma (\alpha+1)}$. This factor represents the contribution of each of the $n-1$ past events on the present event of $n$.

\begin{figure*}{ht}
    \hspace*{-1cm}
    \centering
    \includegraphics[width=12cm, height=6cm]{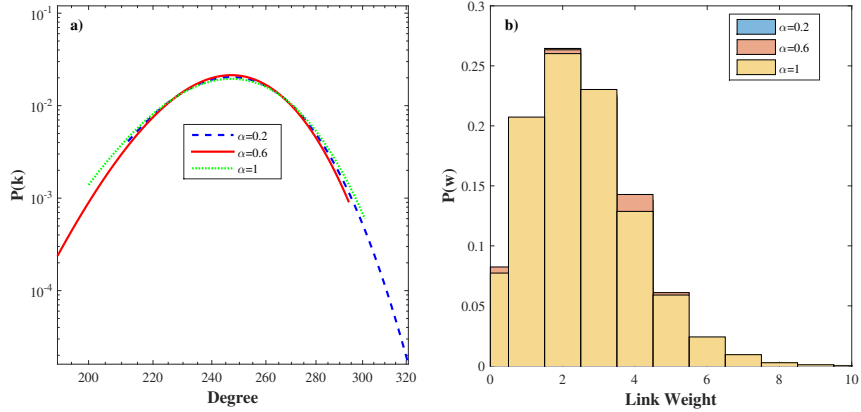}%
    \caption {In this figure, we show the probability distribution function of degree (P(k)) and link weight (P(w)) for different orders of the fractional derivative in time step $t=10^3$. The networks have $N = 200$ nodes with $m=10$, and simulations start with the same random network. For $\alpha =1$ the expected memoryless model behavior is observed, while for fractional orders less than $1$, there is an obvious deviation. This could be caused by two competitive processes, namely a preferential connection mechanism and a memory effect. This is evident from the plots: With the passage of time, the degree distribution function becomes broader.}%
    \label{fig3}%
\end{figure*}

Numerically solving the equation system, Eq.~\eqref{eq:11}, the obtained results confirm the effect of memory on the network evolution. Fig.~\ref{fig2} illustrates the effective node activity for various $\alpha$ values. It is obvious that for all values of $\alpha$, the effective node activity increases rapidly at the beginning and then slows down. However, it is clear that smaller values of $\alpha$, which express the strength of memory in the system, translate into a lower growth of $\bar k_{i}(t)$ compared with higher values of $\alpha$.  In sum, we can explain the initial rapid  increase  of  $\bar k_{i}(t)$  by the fact  that at the  beginning  memory  is  not strong enough  (there is no past)  and nodes connect based on activity; With the  passage  of  time,  memory  becomes  more  important  and  nodes  start  to  prefer the nodes with whom they had connections in the past.
To describe this behavior, it could be helpful to geometrically interpret the meaning of a unit of time in the fractional context. In fractional  space,  time  slows down and expands  at  each  unit,  as  we have shown in Fig.~\ref{fig4}.  At the beginning of fractional time, a unit of time is small compared to the homogeneous timescale, and with time passing, the unit length for fractional time becomes larger than in the homogeneous case. This explains the slowing down of time in Fig.~\ref{fig2}.

In Fig.~\ref{fig3}, we also show some statistical proprieties of the simulated networks. The BA model B predicts that after a transient period the connectivity distribution of all nodes becomes a Gaussian around its mean value. We have shown here that in presence of memory, deviation from a Gaussian occurs, see Fig.~\ref{fig3}-(a). 
This deviation is of course understood as resulting from the competition between  memory effects and the preferential connection mechanism. In the present case, getting older reduces the probability of nodes with old links to be selected, since memory reduces the effective degree of the nodes. This reduces the growth rate of older nodes, whence giving nodes with a smaller degree a higher chance to receive new links. Indeed, Fig.~\ref{fig3}-(a) illustrates that as time increases, the $P(k)$ for $\alpha<1$ shifts from the initial symmetric to a broader .


Memory affects the degree distribution.  For many nodes, memory increases the opportunity to develop into a hub, as opposed to the BA model where only early members have a chance of becoming a hub.  Memory also impacts assortativity measure, defined as the tendency of nodes to connect to those with whom they had interactions in the past. Assortativity increases in networks with aged links. This can be seen in the normal distribution of weights in Fig.~\ref{fig3}-(b).

\section {Conclusion}

Memory can have a great impact on the temporal evolution of networks. In this paper, we rely on fractional calculus to incorporate memory in a temporal network and examine its effect on link formation. We examine two models, the first is a preferential attachment mechanism without memory, and the second includes memory. The results show that the dynamics of temporal and evolving networks depends on the strength of memory effects, controlled by the order of fractional derivatives. We show also that the evolution of a temporal network with incorporated memory depends on the preferential mechanism at the early stage, while memory effects become more relevant at the later stage.

\section{Acknowledgment}
We thank Prof. Rosario N. Mantegna for comments that greatly improved the manuscript.


\begin{thebibliography}{10}
	\expandafter\ifx\csname url\endcsname\relax
	\def\url#1{\texttt{#1}}\fi
	\expandafter\ifx\csname urlprefix\endcsname\relax\def\urlprefix{URL }\fi
	\expandafter\ifx\csname href\endcsname\relax
	\def\href#1#2{#2} \def\path#1{#1}\fi
	
	\bibitem{clauset2009power}
	A.~Clauset, C.~R. Shalizi, M.~E. Newman, Power-law distributions in empirical
	data, SIAM review 51~(4) (2009) 661--703.
	
	\bibitem{newman2011structure}
	M.~Newman, A.-L. Barabasi, D.~J. Watts, The structure and dynamics of networks,
	Vol.~12, Princeton University Press, 2011.
	
	\bibitem{latora2017complex}
	V.~Latora, V.~Nicosia, G.~Russo, Complex networks: principles, methods and
	applications, Cambridge University Press, 2017.
	
	\bibitem{holme2012temporal}
	P.~Holme, J.~Saram{\"a}ki, Temporal networks, Physics reports 519~(3) (2012)
	97--125.
	
	\bibitem{masuda2016guidance}
	N.~Masuda, R.~Lambiotte, A guidance to temporal networks, World Scientific,
	2016.
	
	\bibitem{holme2005network}
	P.~Holme, Network reachability of real-world contact sequences, Physical Review
	E 71~(4) (2005) 046119.
	
	\bibitem{halinen2013network}
	A.~Halinen, J.-{\AA}. T{\"o}rnroos, M.~Elo, Network process analysis: An
	event-based approach to study business network dynamics, Industrial Marketing
	Management 42~(8) (2013) 1213--1222.
	
	\bibitem{li2017fundamental}
	A.~Li, S.~P. Cornelius, Y.-Y. Liu, L.~Wang, A.-L. Barab{\'a}si, The fundamental
	advantages of temporal networks, Science 358~(6366) (2017) 1042--1046.
	
	\bibitem{kostakos2009temporal}
	V.~Kostakos, Temporal graphs, Physica A: Statistical Mechanics and its
	Applications 388~(6) (2009) 1007--1023.
	
	\bibitem{kim2012temporal}
	H.~Kim, R.~Anderson, Temporal node centrality in complex networks, Physical
	Review E 85~(2) (2012) 026107.
	
	\bibitem{wang2017link}
	T.~Wang, X.-S. He, M.-Y. Zhou, Z.-Q. Fu, Link prediction in evolving networks
	based on popularity of nodes, Scientific reports 7~(1) (2017) 7147.
	
	\bibitem{hassanibesheli2017glassy}
	F.~Hassanibesheli, L.~Hedayatifar, H.~Safdari, M.~Ausloos, G.~Jafari, Glassy
	states of aging social networks, Entropy 19~(6) (2017) 246.
	
	\bibitem{kim2015scaling}
	H.~Kim, M.~Ha, H.~Jeong, Scaling properties in time-varying networks with
	memory, The European Physical Journal B 88~(12) (2015) 315.
	
	\bibitem{scholtes2014causality}
	I.~Scholtes, N.~Wider, R.~Pfitzner, A.~Garas, C.~J. Tessone, F.~Schweitzer,
	Causality-driven slow-down and speed-up of diffusion in non-markovian
	temporal networks, Nature communications 5 (2014) 5024.
	
	\bibitem{vestergaard2014memory}
	C.~L. Vestergaard, M.~G{\'e}nois, A.~Barrat, How memory generates heterogeneous
	dynamics in temporal networks, Physical Review E 90~(4) (2014) 042805.
	
	\bibitem{pfitzner2013betweenness}
	R.~Pfitzner, I.~Scholtes, A.~Garas, C.~J. Tessone, F.~Schweitzer, Betweenness
	preference: Quantifying correlations in the topological dynamics of temporal
	networks, Physical review letters 110~(19) (2013) 198701.
	
	\bibitem{karsai2014time}
	M.~Karsai, N.~Perra, A.~Vespignani, Time varying networks and the weakness of
	strong ties, Scientific reports 4 (2014) 4001.
	
	\bibitem{scholtes2013slow}
	I.~Scholtes, F.~Schweitzer, R.~Pfitzner, A.~Garas, N.~Wider, C.~J. Tessone,
	Slow-down vs. speed-up of information diffusion in non-markovian temporal
	networks, Tech. rep. (2013).
	
	\bibitem{karsai2012correlated}
	M.~Karsai, K.~Kaski, J.~Kert{\'e}sz, Correlated dynamics in egocentric
	communication networks, Plos one 7~(7) (2012) e40612.
	
	\bibitem{lambiotte2014effect}
	R.~Lambiotte, V.~Salnikov, M.~Rosvall, Effect of memory on the dynamics of
	random walks on networks, Journal of Complex Networks 3~(2) (2014) 177--188.
	
	\bibitem{williams2019effects}
	O.~E. Williams, F.~Lillo, V.~Latora, Effects of memory on spreading processes
	in non-markovian temporal networks, New Journal of Physics 21~(4) (2019)
	043028.
	
	\bibitem{granovetter1977strength}
	M.~S. Granovetter, The strength of weak ties, in: Social networks, Elsevier,
	1977, pp. 347--367.
	
	\bibitem{dodds2003experimental}
	P.~S. Dodds, R.~Muhamad, D.~J. Watts, An experimental study of search in global
	social networks, science 301~(5634) (2003) 827--829.
	
	\bibitem{onnela2007structure}
	J.-P. Onnela, J.~Saram{\"a}ki, J.~Hyv{\"o}nen, G.~Szab{\'o}, D.~Lazer,
	K.~Kaski, J.~Kert{\'e}sz, A.-L. Barab{\'a}si, Structure and tie strengths in
	mobile communication networks, Proceedings of the national academy of
	sciences 104~(18) (2007) 7332--7336.
	
	\bibitem{song2010modelling}
	C.~Song, T.~Koren, P.~Wang, A.-L. Barab{\'a}si, Modelling the scaling
	properties of human mobility, Nature Physics 6~(10) (2010) 818.
	
	\bibitem{metzler2000random}
	R.~Metzler, J.~Klafter, The random walk's guide to anomalous diffusion: a
	fractional dynamics approach, Physics reports 339~(1) (2000) 1--77.
	
	\bibitem{richard2014fractional}
	H.~Richard, Fractional calculus: an introduction for physicists, World
	Scientific, 2014.
	
	\bibitem{ebadi2016effect}
	H.~Ebadi, M.~Saeedian, M.~Ausloos, G.~R. Jafari, Effect of memory in
	non-markovian boolean networks illustrated with a case study: A cell cycling
	process, EPL (Europhysics Letters) 116~(3) (2016) 30004.
	
	\bibitem{saeedian2017memory}
	M.~Saeedian, M.~Khalighi, N.~Azimi-Tafreshi, G.~R. Jafari, M.~Ausloos, Memory
	effects on epidemic evolution: The susceptible-infected-recovered epidemic
	model, Physical Review E 95~(2) (2017) 022409.
	
	\bibitem{safdari2016fractional}
	H.~Safdari, M.~Z. Kamali, A.~Shirazi, M.~Khalighi, G.~Jafari, M.~Ausloos,
	Fractional dynamics of network growth constrained by aging node interactions,
	PLOS one 11~(5) (2016) e0154983.
	
	\bibitem{berkowitz2002physical}
	B.~Berkowitz, J.~Klafter, R.~Metzler, H.~Scher, Physical pictures of transport
	in heterogeneous media: Advection-dispersion, random-walk, and fractional
	derivative formulations, Water Resources Research 38~(10) (2002) 9--1.
	
	\bibitem{kaslik2012nonlinear}
	E.~Kaslik, S.~Sivasundaram, Nonlinear dynamics and chaos in fractional-order
	neural networks, Neural Networks 32 (2012) 245--256.
	
	\bibitem{west2015fractional}
	B.~J. West, M.~Turalska, P.~Grigolini, Fractional calculus ties the microscopic
	and macroscopic scales of complex network dynamics, New Journal of Physics
	17~(4) (2015) 045009.
	
	\bibitem{Caputo}
	M.~Caputo, \href{https://doi.org/10.1111/j.1365-246X.1967.tb02303.x}{{Linear
			Models of Dissipation whose Q is almost Frequency Independent—II}},
	Geophysical Journal International 13~(5) (1967) 529--539.
	\newblock \href {http://dx.doi.org/10.1111/j.1365-246X.1967.tb02303.x}
	{\path{doi:10.1111/j.1365-246X.1967.tb02303.x}}.
	\newline\urlprefix\url{https://doi.org/10.1111/j.1365-246X.1967.tb02303.x}
	
	\bibitem{podlubny}
	I.~Podlubny, Fractional differential equations: an introduction to fractional
	derivatives, fractional differential equations, to methods of their solution
	and some of their applications, Vol. 198, Elsevier, 1998.
	
	\bibitem{podlubny2001geometric}
	I.~Podlubny, Geometric and physical interpretation of fractional integration
	and fractional differentiation, arXiv preprint math/0110241.
	
	\bibitem{grindrod2009evolving}
	P.~Grindrod, D.~J. Higham, Evolving graphs: dynamical models, inverse problems
	and propagation, Proceedings of the Royal Society A: Mathematical, Physical
	and Engineering Sciences 466~(2115) (2009) 753--770.
	
	\bibitem{buscarino2008disease}
	A.~Buscarino, L.~Fortuna, M.~Frasca, V.~Latora, Disease spreading in
	populations of moving agents, EPL (Europhysics Letters) 82~(3) (2008) 38002.
	
	\bibitem{perra2012activity}
	N.~Perra, B.~Gon{\c{c}}alves, R.~Pastor-Satorras, A.~Vespignani, Activity
	driven modeling of time varying networks, Scientific reports 2 (2012) 469.
	
	\bibitem{starnini2013modeling}
	M.~Starnini, A.~Baronchelli, R.~Pastor-Satorras, Modeling human dynamics of
	face-to-face interaction networks, Physical review letters 110~(16) (2013)
	168701.
	
	\bibitem{barabasi1999emergence}
	A.-L. Barab{\'a}si, R.~Albert, Emergence of scaling in random networks, science
	286~(5439) (1999) 509--512.
	
	\bibitem{barabasi1999mean}
	A.-L. Barab{\'a}si, R.~Albert, H.~Jeong, Mean-field theory for scale-free
	random networks, Physica A: Statistical Mechanics and its Applications
	272~(1-2) (1999) 173--187.
	
	\bibitem{diethelm}
	K.~Diethelm, N.~J. Ford, A.~D. Freed, A predictor-corrector approach for the
	numerical solution of fractional differential equations, Nonlinear Dynamics
	29~(1-4) (2002) 3--22.
	
	\bibitem{garrappa2010linear}
	R.~Garrappa, On linear stability of predictor--corrector algorithms for
	fractional differential equations, International Journal of Computer
	Mathematics 87~(10) (2010) 2281--2290.
	
\end{thebibliography}

\end{document}